# Hidden spin polarization in inversion-symmetric bulk crystals

Xiuwen Zhang[1,2,3,*], Qihang Liu[1,4,*], Jun-Wei Luo[3,#], Arthur. J. Freeman[4] and Alex Zunger[1,#]

[1]University of Colorado, Boulder, CO 80309, USA

[2]Colorado School of Mines, Golden, CO 80401, USA

[3]National Renewable Energy Laboratory, Golden, CO 80401, USA

[4]Deprtment of Physics and Astronomy, Northwestern University, Evanston, IL 60208, USA

[*]These authors contributed equally to this work.

[#]E-mail: jun-wei.luo@nrel.gov; alex.zunger@colorado.edu

Spin-orbit coupling (SOC) can induce spin polarization in nonmagnetic 3D crystals when the inversion symmetry is broken, as manifested by the bulk Rashba (R-1) and Dresselhaus (D-1) effects. We determine that these spin polarization effects originate fundamentally from specific atomic site asymmetries, rather than from the generally accepted asymmetry of the crystal space-group. This understanding leads to the recognition that a previously overlooked hidden form of spin polarization should exist in centrosymmetric materials. Although all energy bands must be doubly degenerate in centrosymmetric materials, we find that the two components of such doubly degenerate bands could have opposite polarizations each spatially localized on one of the two separate sectors forming the inversion partners. We demonstrate such hidden spin polarizations in *centrosymmetric crystals* (denoted as R-2 and D-2) by first-principles calculations. This new understanding could considerably broaden the range of currently useful spintronic materials and enable control of spin polarization via operations on atomic scale.

Much of the recent interest in spintronics[1, 2] has been prompted by the fact that when spin-orbit coupling (SOC) occurs in systems with sufficiently low crystalline symmetry, an effective magnetic field $B_{eff} = \lambda[\nabla V(\boldsymbol{r}) \times \boldsymbol{p}]$ emerges [where $V(\boldsymbol{r})$ denotes the crystal potential and $\boldsymbol{p}$ the momentum])[3] that leads to spin splitting and spin polarization even in nonmagnetic materials. This is known in three-dimensional (3D) bulk-periodic solids as the Dresselhaus effect (denoted here as D-1)[4], arising from "bulk inversion asymmetry" (BIA), and in 2D quantum wells and heterostructures as the Rashba effect[5], arising from 2D "structural inversion asymmetry" (SIA). However, recent theoretical[6,7] and experimental studies[6] revealed a Rashba-like spin-texture in some 3D *bulk crystals* (denoted here R-1), where SIA plays no role, a surprising finding that highlights the incompleteness of our current understanding of the origins of the bulk SOC effects. Indeed, the current view that R-1 and D-1 require absence of crystal bulk inversion symmetry has also established the generally accepted expectation that it would be foolish to look for spin splitting in 3D material that has bulk inversion symmetry, a paradigm that has narrowed enormously the playing field of spintronic materials to non-centrosymmetric materials with significant SOC (large atomic numbers).

The central insight here is that because the SOC is a relativistic effect, anchored on particular nuclear sites in the solid[8], it is the symmetry of such *individual atomic sites* in the solid that forms a good starting point to describe the SOC-induced spin polarization effect, rather than the global symmetry of the unit cell (as in SIA or BIA). The total spin polarization of a crystal is then the (vector) sum over all sites of the local site (Rashba or Dresselhaus) spin polarizations. This simple view not only explains the origin of R-1 and D-1 effects in an unified fashion, but immediately leads to the discovery of two hitherto missing forms of spin polarization in nonmagnetic materials (called here R-2 and D-2), as explained schematically in Figs. 1a, b, c, as follows:

When the site point group of an atom within a *3D* crystal lacks inversion symmetry, the atomic site can either be nonpolar or polar, that is, have either an inversion-asymmetric (IA) local environment, or an local environment creating a site dipole-field (DF)[9], respectively. The atomic SOC of that site will then lead in the former case to *local* Dresselhaus spin polarization, whereas in the latter case it will lead to a *local* Rashba spin polarization (Fig. 1b). When, in addition to these *site* symmetries, the *global* bulk space group lacks inversion symmetry (i.e, being non-centrosymmetric), the inversion-asymmetric local motifs combine to produce the well-known *bulk* Dresselhaus (D-1) effect[4], whereas the site dipole-field local motifs combine to produce the *bulk* Rashba (R-1) effect[5]. The R-1 and D-1 effects are characterized by a *net* spin polarization with their own characteristic spin textures[6, 10, 11]: generally in R-1 we have helical texture [10, 11], where the spin orients perpendicular to dipole field axis[6], and in D-1 non-helical texture.

This insight on how the bulk R-1 and D-1 effects emerge from a superposition of the respective site properties suggests that even when the bulk space group of a solid has inversion symmetry (being centrosymmetric) — a case that would be dismissed as being totally uninteresting from the point of view of spin polarization physics—there can be a *hidden spin polarization* (Fig. 1c). Imagine, for example, a layered crystal with each of its individual layers ("sectors") having a local symmetry producing a dipole field (thus, a local Rashba spin polarization), but the crystal as a whole has a space group with inversion symmetry. In this case, the Rashba spin polarization from each local sector (α) is compensated by the other sector—its inversion partner (β). Rather than being intrinsically absent, as normally expected under the current paradigm, this R-2 spin polarization would instead be concealed by compensation. We illustrate such spin polarization with Dresselhaus or Rashba spin textures in *centrosymmetric crystals* by density functional theory (DFT) calculations: D-2 effect in NaCaBi, and even in bulk Si (diamond-like structure), and R-2 effect in $LaOBiS_2$. The relevant energy bands are doubly degenerate because of a combination of bulk inversion symmetry and time reversal symmetry.

If two quantum mechanical wave functions are degenerate, one can choose their linear combination so the spin polarization of each state is not unique and only the sum of the spin polarization of the two states is meaningful. *A key observation here is that in the R-2 and D-2 effects the spin polarization of the energetically degenerate bands is spatially segregated into a dominant spin texture for one real-space sector, whereas the opposite spin texture is associated with its inversion partner.* This type of R-2 and D-2 doubly degenerate bands is different from the common notion of spin-degenerate bands where both spin-up and spin-down wavefunctions occur in the same real-space sector, i.e., there is no spatially separation between two spin-subbands. Such *local* SOC-induced spin polarization that is segregated into different real space domains is clearly apparent from DFT calculations reported here, because in such calculations the spin degrees of freedom of each bulk state can be projected on individual real-space sectors.

***Bulk symmetry, site symmetry and spin polarization.*** The basic concepts of symmetry used in this paper are illustrated in Fig. 1. One can distinguish the presence or absence of inversion symmetry *in the bulk space group* (centrosymmetric vs. non-centrosymmetric, respectively), from the presence or absence of such inversion symmetry *in the atomic site point group.* The latter is the subgroup formed by the subset of all symmetry operations of the crystal space group that leave that atomic site invariant[12]. The net spin polarization of a bulk crystal is the sum over all atomic sites of the local site spin polarizations. One can then classify all crystal systems into three general spin polarization cases by considering different combinations of bulk space group and site point group. This is explained as cases a, b and c in Fig. 1.

*The three classes of spin polarization in nonmagnetic bulks (Fig. 2).*

*(a) Absence of spin polarization.* When all the site point groups contain inversion symmetry (point groups $C_i$, $C_{2h}$, $D_{2h}$, $C_{4h}$, $D_{4h}$, $S_6$, $D_{3d}$, $C_{6h}$, $D_{6h}$, $T_h$, and $O_h$), there is no SOC-induced *local* spin polarization, and thus the bulk spin polarization vanishes too (Fig. 1a and Fig. 2f). This is the case in the crystal structures such as NaCl-type (*Fm-3m*), CsCl-type (*Pm-3m*), CaTiO$_3$-type perovskite (*Pm-3m*), and Mg-type hexagonal close packed (*P6$_3$/mmc*) structures. For example, the β-phase (NaCl-type) of SnTe—a topological crystalline insulator[13] with *topologically protected states* on surfaces—is found here to have vanishing spin polarization in the *bulk* (Fig. 2f). However, we will show below that such bulk spin polarization can be controlled by tuning the atomic arrangement (thus symmetry), e.g., changing the β-phase having no spin polarization to an α-phase having strong R-1 polarization (Fig. 2c).

*(b) Net bulk spin polarization: The R-1 and D-1 effects.* In these cases, the energy bands have non-zero net spin polarization. We next identify the pertinent combinations of bulk inversion asymmetry and site point groups leading to these macroscopic spin polarization effects (Fig. 2).

*The D-1 effect (Figs. 2a, 2b)* is associated with non-centrosymmetric space group (bulk inversion asymmetry). In this case either all sites have non-polar point group symmetry ($D_2$, $D_3$, $D_4$, $D_6$, $S_4$, $D_{2d}$, $C_{3h}$, $D_{3h}$, $T$, $T_d$, and $O$), or some sites have polar point group symmetry ($C_1$, $C_2$, $C_3$, $C_4$, $C_6$, $C_{1v}$, $C_{2v}$, $C_{3v}$, $C_{4v}$, and $C_{6v}$) and all the site dipoles add up to zero. The simplest example (Fig. 2a) is the zinc-blende structure where the bulk space group is *F-43m* and site point group is $T_d$ for both atom species, such as InAs, GaSb, and GaAs[3]. A similar example is the half-Heusler materials having also bulk space group *F-43m* and site point group $T_d$, which raise a lot of interest for topological insulators[14], such as LaPdBi, LaPtBi, and ZrCoBi. The D-1 effect in these solids leads to spin splitting and as a consequence establishes net bulk spin polarization. As a basis for our comparison below with the new case of D-2, the D-1 behavior is illustrated for ZrCoBi via DFT calculations of the band structure and spin texture in Supplementary Section A.

*The R-1 effect (Fig. 2c)* is accompanied in bulk crystals by the D-1 effect (called together "R-1 & D-1"). They originate from the combination of non-centrosymmetric space group (bulk inversion asymmetry) with the *site dipole field*. Specifically, at least one site must have a polar site point group ($C_1$, $C_2$, $C_3$, $C_4$, $C_6$, $C_{1v}$, $C_{2v}$, $C_{3v}$, $C_{4v}$, and $C_{6v}$) and the individual dipoles add up to a non-zero value (Fig. 2c). Because a polar point group contains simultaneously a site dipole field and site inversion asymmetry (Fig. 2c), the R-1 effect in bulk crystals is always accompanied by the D-1 effect with system-dependent relative magnitude (see below). The R-1 & D-1 effects can occur in the following bulk structure-types: ZnS-type wurtzite (*P6₃mc*), GeTe-type (*R3m*), LiGaGe-type (*P6₃mc*), and BiTeI-type (*P3m1*) structures. Fig. 3 illustrates the R-1 & D-1 effect at both conduction bands (CB1, CB2) and valence bands (VB1, VB2) of α-SnTe[15]. Fig. 3a shows the crystal structure of α-SnTe (space group *R3m*) having polar point groups $C_{3v}$ for both the Sn and Te sites. Thus, this phase belongs to the R-1 & D-1 class in Fig. 2c. The calculated band structure shown in Fig. 3b reveals that R-1 & D-1 effects lift the spin degeneracy between CB1 and CB2 and between VB1 and VB2 creating a significant (~300 meV) spin splitting. The corresponding spin textures obtained from all atoms in the unit cell are shown in Figs. 3c and 3d and exhibit strong spin polarization. Figs. 3e and 3f show more specifically a 2D view of the spin textures of CB1 and VB1 bands, illustrating the *helical in-plane* pattern, reflecting the dominance of R-1 over D-1 in this material. The sensitive dependence of spin-polarization classes on details of the symmetry is clearly demonstrated in the SnTe system: Whereas α-SnTe in the rhombohedral *R3m* bulk symmetry is predicted here to show clear R-1 spin signature (Fig. 2c and Fig. 3), when its rhombohedral distortion (that induces polar point groups on the atomic sites) is removed by deforming the lattice to the centrosymmetric β-phase (total energy rises by ~ 0.01 eV/atom), the spin polarization disappears, as it now belongs to the unpolarized class as illustrated in Fig. 2f.

*(c) Compensated (hidden) spin polarization: The R-2 and D-2 effects.* These can arise in crystal structures where inversion symmetry is present in the bulk space group, but not in the site point groups (Fig. 1c). This is the case when the individual sites either carry a local dipole field (for R-2), or a site inversion asymmetric crystal field (for D-2). A combination of a bulk centrosymmetric space group with site dipole field leads to the bulk R-2 effect, while a combination of a global centrosymmetric space group with site inversion asymmetry results in the bulk D-2 effect (Fig. 1c). We next explain the R-2 and D-2 effects in greater detail.

*The D-2 effect (Fig. 2d)* originates from the combination of centrosymmetric global space group with *site inversion asymmetry*. Specifically, all atoms have non-polar site point groups and, at least one site must have an inversion asymmetric site point group ($D_2$, $D_3$, $D_4$, $D_6$, $S_4$, $D_{2d}$, $C_{3h}$, $D_{3h}$, $T$, $T_d$, and $O$). Such site inversion-asymmetry induced compensated spin polarization (D-2) can occur in a number of structure types, e.g., BN-type ($P6_3/mmc$), ZrBeSi-type ($P6_3/mmc$), and diamond ($Fd$-$3m$) structures. Fig. 4 illustrates the D-2 effect in NaCaBi having the ZrBeSi-type structure with space group $P6_3/mmc$ and non-centrosymmetric site point group $D_{3h}$ for both Ca and Bi atoms, as well as centrosymmetric site point group $D_{3d}$ for Na atom (Fig. 4a). The two CaBi layers connected by inversion symmetry are indicated in Fig. 4a as α-sector and β-sector. The calculated band structure as shown in Fig. 4b illustrates clearly large band splitting (~400 meV) in the vicinity of $L$ point in the Brillouin zone between CB1 and CB2 and between VB1 and VB2. The corresponding spin texture, projected on the real-space sectors α and β, forming the inversion partners, are shown as brown and green arrows, respectively in Figs. 4c and 4d. Even though all the bands are energetically doubly degenerate due to bulk inversion symmetry, we see that *each of the branches of the two-fold degenerate band has opposite spin polarization in real space*. This real-space separation originates from the separate local site Dresselhaus SOC, as indicated by the brown branch projected on the α-sector (CaBi)$_α$ layer and the green one projected on the β-sector (CaBi)$_β$ layer.

Centrosymmetric *bulk* silicon was always assumed to be a spin polarization-free material (so it is assumed not to disturb externally injected spin polarization[16]). It is an ideal material for long spin lifetime since the D-1 effect is absent and the effect of nuclear spin of Si vanishes[17]. However, we illustrated in Supplementary Section B that, bulk diamond-like silicon has compensated D-2 spin polarization (the class shown in Fig. 2d) because the site point group of Si atoms is non-centrosymmetric $T_d$ which leads to local atomic Dresselhaus spin polarization.

*The R-2 effect (Fig. 2e)* is accompanied in bulk crystals by the D-2 effect (called together "R-2 & D-2"). They originate from the combination of centrosymmetric bulk space group with the "site dipole field". Specifically, at least one site must have a polar site point group ($C_1$, $C_2$, $C_3$, $C_4$, $C_6$, $C_{1v}$, $C_{2v}$, $C_{3v}$, $C_{4v}$, and $C_{6v}$, see Fig. 2e). The R-2 & D-2 spin polarization can occur in structure types such as MoS$_2$-type (*P6$_3$/mmc*), Bi$_2$Te$_3$-type (*R-3m*), and LaOBiS$_2$-type (*P4/nmm*) structures. Fig. 5 illustrates the R-2 & D-2 effects at CB1, CB2, VB1 and VB2 in LaOBiS$_2$ which is recently discovered to have similar properties with cuprate- and iron-based superconductors[18]. Fig. 5a shows the crystal structure of LaOBiS$_2$ having a centrosymmetric bulk space group and non-centrosymmetric polar site point group $C_{4v}$ for Bi, S, and La atoms, as well as non-centrosymmetric non-polar site point group $S_4$ for O atoms. The unit cell of LaOBiS$_2$ can be divided into three sectors: two inversion-partners BiS$_2$ layers termed α-sector and β-sector in Fig. 5a and the central La$_2$O$_2$ layer that does not affect the low-energy (near band gap) spectrum. The calculated band structure as shown in Fig. 5b illustrates clearly large band splitting (~120 meV) in the vicinity of $X$ point in the Brillouin zone between CB1 and CB2 and between VB1 and VB2. These are induced mainly by the large site dipole field as well as by the site inversion asymmetry in the BiS$_2$ layers. The corresponding spin textures projected on sectors α and β are shown in Figs. 5c and 5d.

They manifest two branches of spin polarization (indicated in Figs. 5c and 5d by green and brown arrows) corresponding to the two $BiS_2$ real-space sectors shown in Fig. 5a. We see in Figs. 5c-5d that the spin polarizations from the two inversion-symmetric $BiS_2$ layers have opposite directions and compensate each other. The 2D spin textures of CB1 and VB1 are shown in Figs. 5e and 5f. We see both helical (mainly in Fig. 5f) and non-helical (mainly in Fig. 5e) spin textures, suggesting that R-2 effect is accompanied by D-2 effect. The distribution of R-2 and D-2 effects on each band depends on the specific band character.

**Discussion** *(i) Physical generality of the observation*: The concept that opposing local effects can mutually compensate in an extended crystal is common to the present discussion on spin polarization as well as to the much earlier work of Vaida et al. on pyroelectricity[19] who discusses classic electro-elasticity and compensation by molecular units ("sectors"). In the present study we point out that the polar/nonpolar *point group symmetry of sites* carries the essential information. The connection to point group symmetry allows us to accomplish two new results: (a) identify other properties previously thought to exist only in non-centrosymmetric crystals (e.g., spin polarization, pyroelectricity, and second harmonic generation) as belonging to this general class of *hidden effects in centrosymmetric crystal structures*. This is demonstrated in Fig. 2 and Supplementary Section C that show how these three types of observables emerge in specific combinations of site point groups and space groups. (b) The focus on site symmetries further allows us to systematically predict actual *materials* that would show such hidden compensation effects, for example, NaCaBi shows the D-2 effect whereas $LaOBiS_2$ shows the R-2 effect, as described in what follows.

*(ii) How can one find actual materials with the R-2 or D-2 effect*: Our objective is to transform the understanding of the symmetry underpinning of the R-1, D-1, R-2, and D-2 effects to a search algorithm of actual materials that will manifest these effects, and then verify our understanding by calculating the respective spin texture for such prototypical materials. This is done by the following design principles: (a) Look for compounds that have centrosymmetric space groups but at least one of the Wyckoff positions lacks inversion symmetry and belongs to either polar (for R-2) or nonpolar (for D-2) point groups (listed in Fig. 2). Out of 230 space groups, 92 are centrosymmetric; out of 32 point groups, 21 are non-centrosymmetric[20]. (b) To find materials that have a *significant* R-2 or D-2 spin polarization, look for the compounds that have heavy elements located on atomic sites with *a large dipole field*; the *latter* can be readily obtained from an electronic structure calculation within first principles DFT by computing the gradient of the self-consistent electrostatic potential at various lattice sites. For $LaOBiS_2$ we find a rather large dipole field of $1.8 \times 10^5$ KV/cm on Bi site (see details in Supplementary Section D) (c) We find (Supplementary Section E) that *layered* materials having a *small* interaction between the layers enhance the polarization effect, as this acts to avoid the tendency for cancellation between the opposite spin polarizations from the respective layers. This is the case in $LaOBiS_2$ having a $La_2O_2$ barrier separating the two $BiS_2$ layers. As a final validation step, one can calculate the spin polarization and splitting as in Figs. 4 and 5. The choices of the illustrative compounds in this paper, i.e., $LaOBiS_2$ and NaCaBi, are all inspired by these design principles, and thus their calculated spin polarizations and splittings are quite large. Therefore, our approach is not only descriptive, but it is also predictive.

*(iii) Detectability*: The real-space segregation of spin polarization in R-2 or D-2 holds the potential for detectability of such hidden spin polarization. ***Bulk*** spin polarization in such centrosymmetric layered materials could be observed when a probing beam penetrating this material along the α/β/α/β... stacking direction is attenuated with depth, so the spin polarization of layer α is not exactly compensated by the spin polarization of its inversion partner β (see a simple description in Supplementary Section E). This bulk spin effect can be distinguished from potential ***surface spin*** effect since the former sensitively depends on the effective penetrating depth of the probing beam[21] while the latter does not. This perspective opens the possibility of *deliberate design* of imperfect compensation from different sectors to maximize spin polarization in centrosymmetric crystals.

*(iv) Potential advantages of compensated over non-compensated spin polarization:* In addition to the obvious increase in the number of systems that can be amenable to spin-polarization based applications, the currently predicted spin polarization in centrosymmetric materials (the R-2 or D-2 effect) might have some advantage over the more conventional R-1 or D-1 effect in non-centrosymmetric systems. Indeed, in the former case one could sensitively manipulate the effect via application of symmetry-breaking external perturbations such as external electric field[22]. In contrast, the more ordinary bulk Rashba effect in non-centrosymmetric crystals (R-1) is difficult to manipulate or reverse by external electric field since the internal field (e.g., $0.4 \times 10^5$ KV/cm on Sn site in α-SnTe, see Supplementary Section D) is usually much stronger than the former. Thus, the hidden spin polarization may provide a new route to operate the electron's spin.

## Methods

In this work we evaluate the band structure by Density Functional Theory (DFT)[23] using the Projector-Augmented Wave (PAW) pseudopotentials[24] with the exchange-correlation of Perdew-Burke-Ernzerhof (PBE) form[25] as implemented in the Vienna Ab-initio Simulation Package (VASP)[26]. We choose the energy-cutoff in the range of 300-550 eV, and the reciprocal space grids of 10×10×10, 18×18×6, 8×8×6, 12×12×12, and 13×13×4 for ZrCoBi, α-SnTe, NaCaBi, Si, and LaOBiS$_2$, respectively. The crystal structure is taken from Ref. [20, 27, 28] and the atomic position is fully relaxed under the tolerance of $10^{-4}$ eV/Å. Spin-orbit coupling is taken into account all though calculated by a perturbation $\sum_{i,l,m} V_l^{SO} \boldsymbol{L} \cdot \boldsymbol{S} |l,m\rangle_{i\ i}\langle l,m|$ to the pseudopotential ($|l,m\rangle_i$ is the angular momentum eigenstate of atomic site $i$)[29]. The spin polarization is evaluated by projecting the calculated wavefunction $|\phi\rangle$ on spin and orbital basis of each atomic site $C_{i,l,m,\eta} = \langle \phi | (s_\eta \otimes |l,m\rangle_{i\ i}\langle l,m|) |\phi\rangle$ and then sum $C_{i,l,m,\eta}$ for a given spin direction and sector that contains a number of atomic sites in the unit cell. The Wigner-Seitz radii for constructing $|l,m\rangle_i$ used in this study are listed in the pseudopotentials of the VASP simulation package[26]. In future, we will use advanced numeric method to project the wavefunction onto any shape of volume of the sector.

Requests for materials should be addressed to A.Z.

## Acknowledgements

A.Z. is grateful to Emmanuel Rashba for important discussions on the manuscript and to Meir Lahav for discussing the analogy with anti-pyroelectricity (Ref. 19). This work was supported by NSF Grant No. DMREF-13-34170. J.-W.L. was supported by the Center for Inverse Design, an Energy Frontier Research Center funded by the US Department of Energy, Office of Science,

## Author Contributions

X.Z. and Q.L. carried out the electronic structure calculations. A.Z. led the analysis and the writing of the paper. J.-W.L. contributed equally with Q.L. and X.Z. to the preparation of the figures and writing of the paper. A.J.F. and A.Z. supervised the study.

## References

1. Wolf, S. A., Awschalom, D. D., Buhrman, R. A., Daughton, J. M., von Molnár, S., Roukes, M. L.*, et al.* Spintronics: A Spin-Based Electronics Vision for the Future. *Science* **294**, 1488-1495 (2001).

2. Žutić, I., Fabian, J. & Das Sarma, S. Spintronics: Fundamentals and applications. *Rev. Mod. Phys.* **76**, 323-410 (2004).

3. Luo, J.-W., Bester, G. & Zunger, A. Full-Zone Spin Splitting for Electrons and Holes in Bulk GaAs and GaSb. *Phys. Rev. Lett.* **102**, 056405 (2009).

4. Dresselhaus, G. Spin-orbit coupling effects in zinc blende structures. *Phys. Rev.* **100**, 580-586 (1955).

5. Rashba, E. I. Properties of semiconductors with an extremum loop. 1. Cyclotron and combinational resonance in a magnetic field perpendicular to the plane of the loop. *Soviet Physics-Solid State* **2**, 1109-1122 (1960).

6. Ishizaka, K., Bahramy, M. S., Murakawa, H., Sakano, M., Shimojima, T., Sonobe, T.*, et al.* Giant Rashba-type spin splitting in bulk BiTeI. *Nat. Mater.* **10**, 521-526 (2011).

7. Di Sante, D., Barone, P., Bertacco, R. & Picozzi, S. Electric Control of the Giant Rashba Effect in Bulk GeTe. *Adv. Mater.* **25**, 509-513 (2013).

8. Herman, F., Kuglin, C. D., Cuff, K. F. & Kortum, R. L. Relativistic Corrections to the Band Structure of Tetrahedrally Bonded Semiconductors. *Phys. Rev. Lett.* **11**, 541-545 (1963).

9. Tilley, R. *Crystals and Crystal Structures*, 2006.

10. Winkler, R. Spin orientation and spin precession in inversion-asymmetric quasi-two-dimensional electron systems. *Phys. Rev. B* **69**, 045317 (2004).

11. Winkler, R. *Spin-Orbit Coupling Effects in Two-Dimensional Electron and Hole Systems*. Springer, 2003.

12. Flurry, R. L. Site symmetry in molecular point groups. *International Journal of Quantum Chemistry* **6**, 455-458 (1972).

13. Xu, S.-Y., Liu, C., Alidoust, N., Neupane, M., Qian, D., Belopolski, I*., et al.* Observation of a topological crystalline insulator phase and topological phase transition in Pb1−xSnxTe. *Nat. Commun.* **3**, 1192 (2012).

14. Chadov, S., Qi, X., Kübler, J., Fecher, G. H., Felser, C. & Zhang, S. C. Tunable multifunctional topological insulators in ternary Heusler compounds. *Nat. Mater.* **9**, 541-545 (2010).

15. Littlewood, P. B. Phase transitions and optical properties of IV-VI compounds. *Lect. Notes Phys.*, vol. 152, 1982, pp 238-246.
16. Dash, S. P., Sharma, S., Patel, R. S., de Jong, M. P. & Jansen, R. Electrical creation of spin polarization in silicon at room temperature. *Nature* **462**, 491-494 (2009).

17. Zhang, L., Luo, J.-W., Saraiva, A., Koiller, B. & Zunger, A. Genetic design of enhanced valley splitting towards a spin qubit in silicon. *Nat. Commun.* **4**, (2013).

18. Mizuguchi, Y., Demura, S., Deguchi, K., Takano, Y., Fujihisa, H., Gotoh, Y*., et al.* Superconductivity in Novel BiS2-Based Layered Superconductor LaO1-xFxBiS2. *J. Phys. Soc. Jap.* **81**, (2012).

19. Vaida, M., Shimon, L. J. W., Weisinger-Lewin, Y., Frolow, F., Lahav, M., Leiserowitz, L*., et al.* The Structure and Symmetry of Crystalline Solid Solutions: A General Revision. *Science* **241**, 1475-1479 (1988).

20. ICSD. Inorganic Crystal Structure Database. *Fachinformationszentrum Karlsruhe: Karlsruhe, Germany*, (2006).

21. Cao, Y., Waugh, J. A., Zhang, X. W., Luo, J. W., Wang, Q., Reber, T. J*., et al.* Mapping the orbital wavefunction of the surface states in three-dimensional topological insulators. *Nat. Phys.* **9**, 499-504 (2013).

22. Liu, Q., Guo, Y. & Freeman, A. J. Tunable Rashba Effect in Two-Dimensional LaOBiS2 Films: Ultrathin Candidates for Spin Field Effect Transistors. *Nano Letters* **13**, 5264-5270 (2013).

23. Kohn, W. & Sham, L. J. Self-Consistent Equations Including Exchange and Correlation Effects. *Phys. Rev.* **140**, A1133-A1138 (1965).

24. Kresse, G. & Joubert, D. From ultrasoft pseudopotentials to the projector augmented-wave method. *Phys. Rev. B* **59**, 1758-1775 (1999).

25. Perdew, J. P., Burke, K. & Ernzerhof, M. Generalized Gradient Approximation Made Simple. *Phys. Rev. Lett.* **77**, 3865-3868 (1996).

26. Kresse, G. & Furthmüller, J. Efficiency of ab-initio total energy calculations for metals and semiconductors using a plane-wave basis set. *Comp. Mater. Sci.* **6**, 15-50 (1996).

27. Zhang, X., Yu, L., Zakutayev, A. & Zunger, A. Sorting Stable versus Unstable Hypothetical Compounds: The Case of Multi-Functional ABX Half-Heusler Filled Tetrahedral Structures. *Adv. Func. Mater.* **22**, 1425-1435 (2012).

28. Tanryverdiev, V. S., Aliev, O. M. & Aliev, I. I. Synthesis and physicochemical properties of LnBiOS{sub 2}. *Inorg. Mater.* **31**, 1497 (1995).

29. Błoński, P. & Hafner, J. Magnetic anisotropy of transition-metal dimers: Density functional calculations. *Phys. Rev. B* **79**, 224418 (2009).

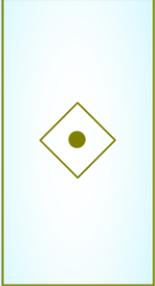

**Figure 1 | The three classes of spin polarization in nonmagnetic bulk crystals.** **a,** *Absence of spin polarization* in centrosymmetric crystals if all atomic sites are inversion symmetric. Since the local environment (crystal field) of centrosymmetric atomic sites does not produce SOC-induced spin polarization, the total (bulk) spin polarization is absent as well. **b,** Net *bulk spin polarization*s (R-1 and D-1 effects): A local site dipole field (DF) or the site inversion asymmetry (IA) leads to a local Rashba or local Dresselhaus effect, respectively. In combination with a ***non-centrosymmetric*** space group, these local effects produce bulk R-1 (Rashba) and D-1 (Dresselhaus) effects, respectively. **c,** *Compensated (hidden) bulk spin polarization* (R-2 and D-2 effects): A local site dipole field (DF) or the site inversion asymmetry (IA) leads to a local Rashba or local Dresselhaus effect, respectively, as in **b**. In combination with a ***centrosymmetric*** space group, these local effects produce bulk R-2 (Rashba) and D-2 (Dresselhaus) effects, respectively. Here the spin polarization from each sector is concealed by compensation from their inversion partners, but are readily visible when the results from individual sectors are observed.

| Site point group / Bulk space group | Non-Centrosymmetric (at least one site) | | | Centrosymmetric (all sites) |
|---|---|---|---|---|
| | Non-polar (all sites) ($D_2$, $D_3$, $D_4$, $D_6$, $S_4$, $D_{2d}$, $C_{3h}$, $D_{3h}$, $T$, $T_d$, $O$) | Polar (at least one site) ($C_1$, $C_2$, $C_3$, $C_4$, $C_6$, $C_{1v}$, $C_{2v}$, $C_{3v}$, $C_{4v}$, $C_{6v}$) | | ($C_i$, $C_{2h}$, $D_{2h}$, $C_{4h}$ $D_{4h}$, $S_6$, $D_{3d}$, $C_{6h}$ $D_{6h}$, $T_h$, $O_h$) |
| | | Dipole adds up to zero | Dipole adds up to non-zero | |
| Non-CS (=BIA) (e.g. F-43m) | **a** D-1 (Dresselhaus) Example: GaAs, ZrCoBi | **b** D-1 Example: γ-LiAlO₂ | **c** R-1 & D-1 Example: BiTeI, α-SnTe | *Not possible (Site point group cannot be CS if space group is non-CS)* |
| Centrosymmetric (e.g. R-3m) | **d** D-2 Example: Si, NaCaBi | **e** R-2 & D-2 Example: MoS₂, Bi₂Se₃, LaOBiS₂ | | **f** Absence of spin polarization Example: β-SnTe |

**Figure 2 | Classification of spin polarization in nonmagnetic bulk materials** based on bulk space group and site point group. The point groups are given in Schoenflies notation. Site point group (SPG) refers to the subset of symmetry operations of bulk space group which transform the atomic site into itself[12]. Polar point groups are the point groups containing a unique anisotropic axis—this causes a net dipole field[9]. CS: centrosymmetric (in γ-LiAlO₂, there are dipole field induced local spin polarizations that are "compensated" by each other, while the D-1 effect induces net spin polarization, i.e. the total spin polarization is not compensated. Thus, we do not claim compensated spin polarization in this material). When the space group is non-CS, at least one site has non-CS point group, thus there is Dresselhaus (D-1) spin polarization (**a**,**b**,**c**). If some of site point groups are polar and the induced dipole fields add up to non-zero, there is a Rashba (R-1) spin polarization (**c**). When the space group is CS, the site point groups can be: (i) all CS in which case (**f**) the spin polarization is absent; (ii) all non-polar but some are non-CS (**d**) in which case there is a D-2 compensated spin polarization, or (iii) containing polar point groups (**e**) in which case there are R-2 accompanied by D-2 compensated spin polarizations.

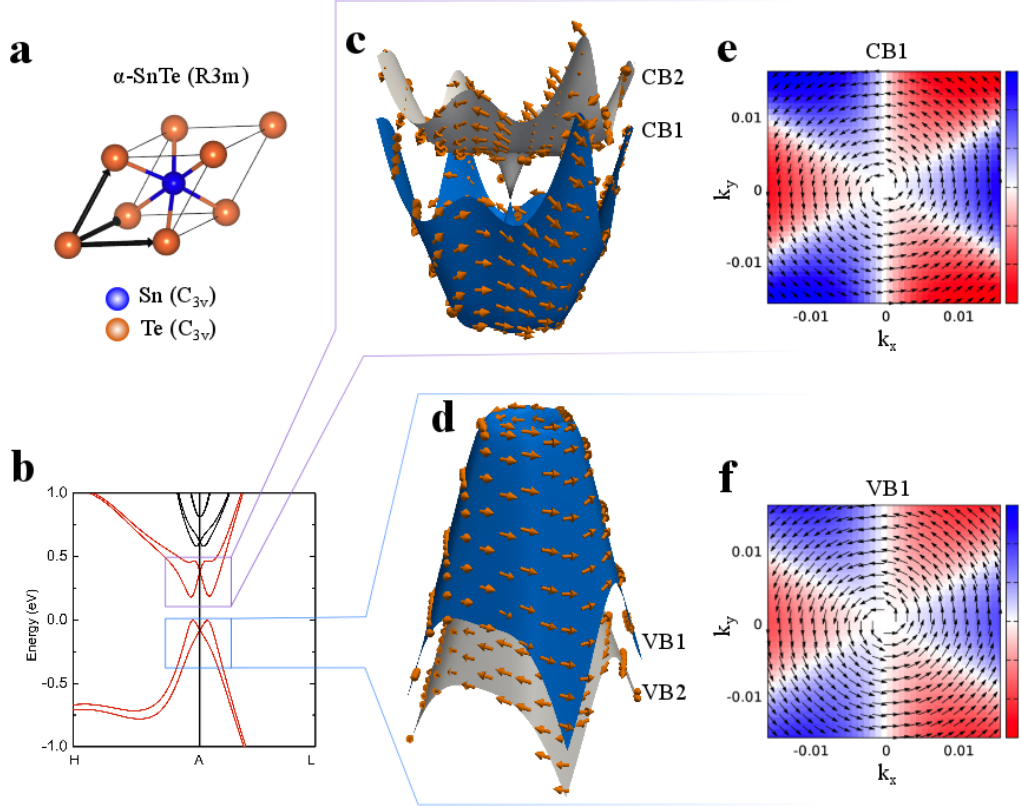

**Figure 3 | Rhombohedral α-SnTe (*R3m*) with R-1 (dominant over D-1) spin textures.**
**a**, Crystal structure and the site point groups of the Sn and Te two atoms. **b**, Band structure around *A* point along the *L-A-H* directions in the Brillouin zone. The two lowest conduction (CB1 and CB2) and two highest valence bands (VB1 and VB2) are shown in red. **c**, **d**, Spin polarization (contributed by the entire unit cell) indicated by brown arrows for the states of spin split CB1 + CB2 and VB1 + VB2 bands in a k-plane, containing *L-A*, *A-H* directions, near *A* point indicated by solid boxes in **b**, respectively. **e**, **f**, Corresponding 2D diagram of spin polarization of CB1 and VB1, respectively. The arrows indicate the in-plane spin direction and the color scheme indicates the out-of-plane spin component. The magnitude of each spin vector is renormalized to 1 for simplicity unless otherwise noted, and notices that the magnitude of spin vectors varies with k-points and bands depending on the strength of spin-orbit coupling and the magnitude of spin splitting.

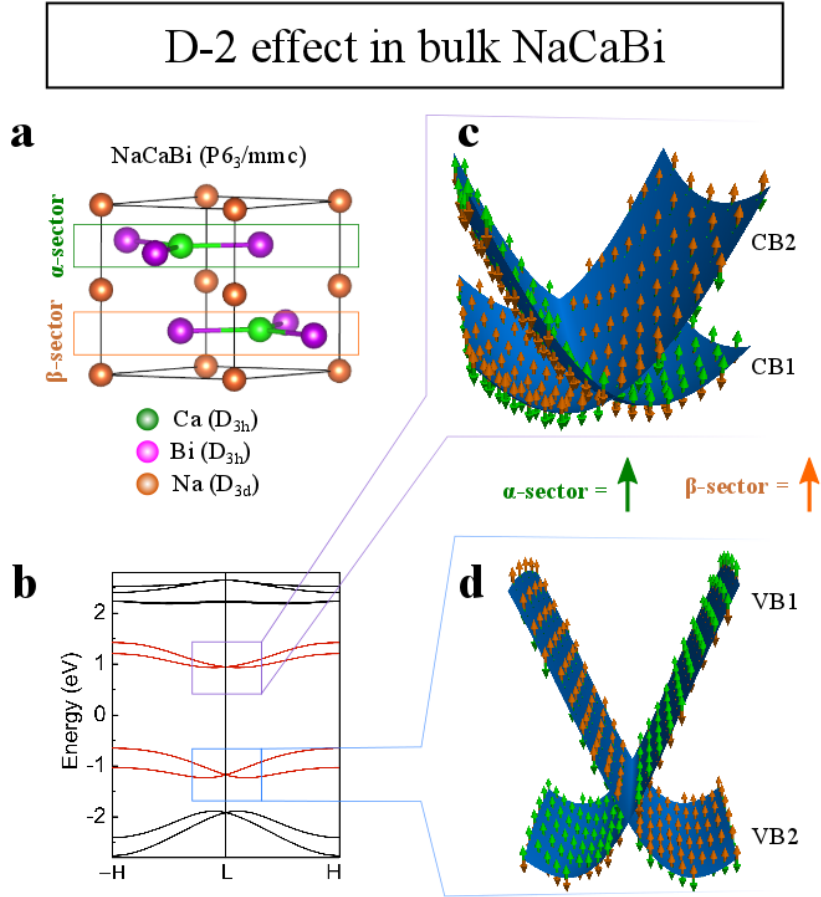

**Figure 4 | NaCaBi (*P6₃/mmc*) with D-2 effect. a**, Crystal structure and the site point group of each atomic site. The two boxed real-space sectors (CaBi layers), forming the inversion partners, used for spin projection are labeled α-sector and β-sector. **b**, Band structure near *L* point along the *L-(-H)* and *L-H* directions in the Brillouin zone. The CB1-CB2 and VB1-VB2 band pairs are shown in red. **c, d**, Projected local spin polarization is represented by green (on α-sector) and brown (on β-sector) arrows for CB1 and CB2, VB1 and VB2 bands in a k-plane, containing *L-H* and *L-A* directions, near *L* point indicated by solid boxes in **b**, respectively. The band crossing at the *L* point is not due to time-reversal symmetry, but due to fractional translation operations, i.e. screw axis *6₃* and glide plane *c* along [0001] direction. We illustrate the compensated spin polarization near the band crossing *L* point just for simplicity and notices that band crossing is not necessary for compensated spin polarization[20].

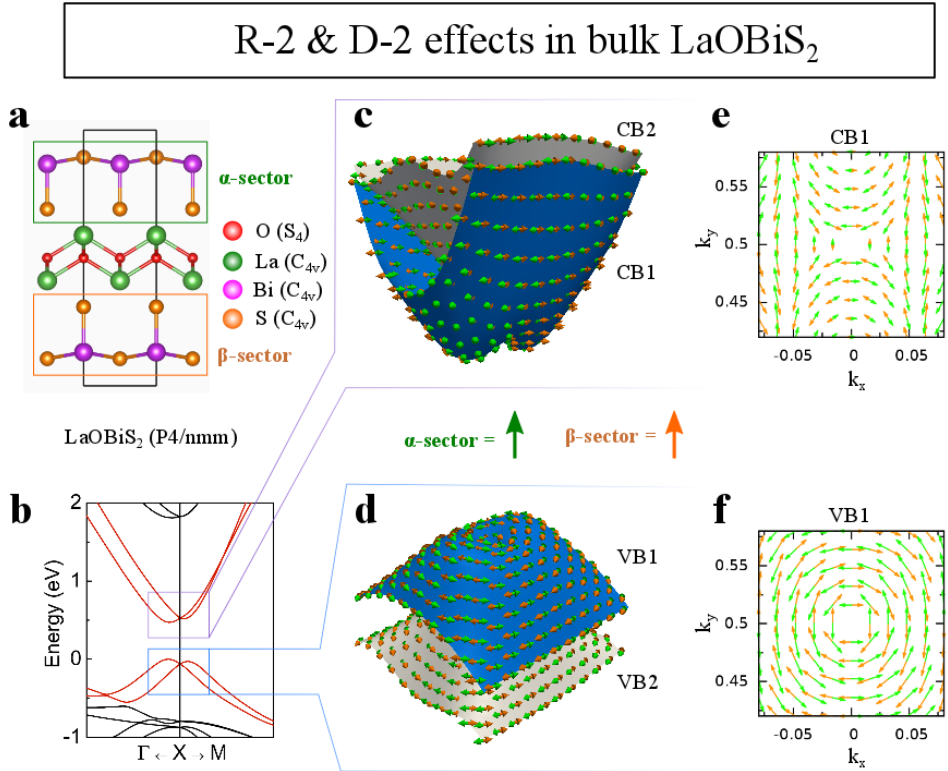

**Figure 5 | LaOBiS₂ (*P4/nmm*) with R-2 & D-2 effects. a**, Crystal structure and the site point group of each atomic site. The two boxed real-space sectors (BiS₂ layers), forming the inversion partners, used for spin projection are labeled α-sector and β-sector. **b**, Band structure around *X* point along *X-Γ* and *X-M* directions in the Brillouin zone. The CB1-CB2 and VB1-VB2 band pairs are highlighted in red. **c, d**, Projected local spin polarization is represented by green (on α-sector) and brown (on β-sector) arrows for CB1 and CB2, VB1 and VB2 bands in the k-plane containing *X-Γ* and *X-M* directions, near *X* point indicated by solid boxes in **b**. **e, f**, 2D diagram of spin polarization of CB1 and VB1, respectively. The arrows indicate the in-plane spin direction. We do not see out-of-plane spin component in **c-f** to the extent of our numeric accuracy. The band crossing at *X* point is not due to time-reversal symmetry, but due to fractional translation operations, i.e. glide plane *n* along [110] direction. We illustrate the compensated spin

polarization near the band crossing $X$ point for simplicity and notices that band crossing is not necessary for compensated spin polarization[20].